\documentstyle[12pt,special,psfig,a4]{article}

\begin{document}
\def\tast{t^\ast}
\def\Jast{J^\ast}
\title{Magnetic Properties of the $t$-$J$ Model in the Dynamical Mean-Field
Theory}

\author{Th.\ Pruschke, Q.\ Qin\cite{qi}, Th.\ Obermeier and
J.\ Keller\\
Institut f\"ur Theoretische Physik, Universit\"at Regensburg,\\ D-93040
Regensburg, Germany
}
\maketitle
\date{\today}
\begin{abstract}
We present a theory for the spin correlation function
of the $t$-$J$ model in the framework of the dynamical mean-field
theory.
Using this mapping between the lattice and a local model we are able to obtain
an intuitive expression for the non-local spin susceptibility, with the
corresponding local correlation function as input.
The latter is calculated by means of local Goldstone diagrams
following closely the procedures developed and successfully applied for the
(single impurity) Anderson model.
We present a systematic study of the magnetic susceptibility and compare our
results with those of a Hubbard model at large $U$. Similarities and
differences
are pointed out and the magnetic phase diagram of the $t$-$J$ model is
discussed.
\end{abstract}
%
Pacs numbers: 71.27+a, 71.28+d, 75.10.LP
%
%
%
%
%
\newpage

\section{Introduction and survey.}
The description of strongly correlated electron systems involves by and
large three different classes of models. First one may consider a system
consisting of uncorrelated delocalized electronic states
hybridizing with localized states subject to a strong Coulomb repulsion.
This situation is modeled by the well known periodic Anderson model \cite{pam}
frequently used to describe the so-called heavy-fermion compounds \cite{steg}.
The second important situation occurs when the delocalized states themselves
feel locally such a strong repulsion. In that case one is led to the
single-band Hubbard model \cite{hm}, originally set up to describe (ferro-)
magnetism and metal-insulator transitions in $3d$ transition-metals compounds
like V$_2$O$_3$ but recently also used for the high-T$_c$ superconductors.
Another interesting kind of system is obtained if in addition to those local
correlations a nonlocal magnetic exchange is included. This is the domain of
the
so-called $t$-$J$ model \cite{tj} which is frequently taken as an alternative
to the Hubbard model to describe the properties of the cuprate superconductors.
It is this model we want to study more closely in this
paper. Although the $t$-$J$ model may be viewed as an effective Hamiltonian for
the
low-energy properties of the Hubbard model in the limit of large local Coulomb
energy \cite{gross87}, i.e.\ vanishing effective magnetic exchange, both
models are expected to differ fundamentally for increasing exchange
interaction.

The Hamiltonian of the $t$-$J$ model reads
\begin{equation}\label{tjm}
H_{t-J}=-\frac{\tast}{\sqrt{2\cal Z}}\suml_{\langle ij\rangle\sigma}
X^{(i)}_{1\sigma,0}X^{(j)}_{0,1\sigma}
+\frac{\Jast}{\cal Z}\suml_{\langle ij\rangle}\vec{S}^i\cdot\vec{S}^j\;\;.
\end{equation}
In equation (\ref{tjm}), $X^{(j)}_{MM'}=|j,M\rangle\langle j,M'|$ are the
standard Hubbard operators \cite{hub2} acting on states with quantum number
$M\in\{0,1\sigma\}$ on site $j$, i.e.\ double occupancy
of a site is strictly forbidden, and $\vec{S}^i$ denotes the spin operator
on site $i$. The sums in the Hamiltonian (\ref{tjm}) are on nearest neighbors
only. The transfer and exchange integrals $t$ and $J$ have been rescaled with
the
coordination number $\cal Z$ of the system
to guarantee a physical meaningful result for large spatial dimensions to
be introduced later. Note that for $\Jast=0$ the model (\ref{tjm}) is the
Hubbard model in the limit $U=\infty$. An additional density-density
interaction
frequently included in the model (\ref{tjm}) has been dropped here for
reasons of convenience.

Although the model (\ref{tjm}) looks rather simple, relatively little is known
{\em exactly}
about its properties. In contrast to the Hubbard model, it is not even exactly
solvable in $d=1$ except for the two special points $\Jast=0$ (Hubbard model)
\cite{liebwu} and $\Jast=2\tast$ (supersymmetric $t$-$J$ model)
\cite{kawakami}. Nevertheless,
exact diagonalization studies showed that the $t$-$J$ model for $d=1$ and $T=0$
is a
Luttinger liquid for all $J<J^{PS}$, while for $J>J^{PS}$ one finds phase
separation into an electron and hole rich region \cite{ogata}. Interestingly,
close to this boundary, the ground state of the $t$-$J$ model is dominated
by superconducting pair correlations \cite{prsh}, while for smaller $J$
antiferromagnetic correlations
are strongest.

Obviously, this would make the $t$-$J$ model an interesting candidate for
explaining e.g.\ high-temperature superconductivity. Unfortunately, the
results for $d=1$ suggest a much too large value of $\Jast/\tast\sim 3\ldots4$
for this scenario.
The interesting question thus is how these features survive in $d>1$ and
especially to what extent phase separation might occur at much lower values
of $J$, as suggested by e.g.\ high-temperature expansions \cite{puttica}.

While in $d=1$ the combination of exact diagonalization and tools of conformal
field theory provides a powerful framework to extract informations about  the
asymptotics of the macroscopic system, similar methods do not exist in $d>1$.
Quantum Monte Carlo techniques, too, cannot be applied for
realistic lattice-sizes and temperatures due to a severe minus-sign problem.
Thus most informations about the properties of the $t$-$J$ model
come from high-temperature expansions, which are restricted to
relatively large values of $\Jast$ and $T$ \cite{puttica,dagotto}, and exact
diagonalization studies for small two-dimensional systems
\cite{dagotto,prelovsek}.
The finite system size in the latter method possibly prevents one from
resolving
dynamically generated low-energy features, which one may especially expect
close
to half filling \cite{pjf,preuss,to}. Moreover,
to interpret results for dynamic quantities calculated with this method one
generally needs additional information from other techniques about the general
structures to be expected.
Clearly, a different approach to obtain results in the thermodynamic limit is
needed.

Usually, a mean-field theory provides a reliable tool to study at least
the qualitative features of models in theoretical solid-state physics.
However, until recently a thermodynamically consistent mean-field theory
like for spin systems did not exist for fermionic models like the $t$-$J$ model
(\ref{tjm}):
While the magnetic exchange term could in principle be handled
by the standard Hartree factorization it is a priori not obvious how to treat
the correlated
hopping introduced by the first term in the model (\ref{tjm}) consistently
within this ansatz. Different schemes, usually involving slave-boson
techniques,
have been proposed \cite{dagotto}. These methods treat the local
dynamics induced by the correlations rather poorly and a systematic inclusion
of fluctuations around the static limit to incorporate lifetime effects is
very cumbersome and has not been successful yet \cite{arigoni}.

Over the past three years, however, a novel scheme was introduced to define
a thermodynamically consistent mean-field theory for correlated systems  that
preserves the local dynamics exactly  \cite{bm,janis,Jarrell}.
In this contribution we shall use this so-called ``dynamical mean-field
theory''
to study the mean-field magnetic properties of the $t$-$J$ model (\ref{tjm}).
The paper is organized as follows. In the next section
we will briefly introduce the dy\-na\-mi\-cal mean-field theory and derive
expressions for the magnetic susceptibility of the $t$-$J$ model. We then
present
results on the magnetic properties and compare them to the large-$U$ Hubbard
model. A summary and discussion concludes the paper.
\section{Theoretical background}
Since the pioneering work of Metzner and Vollhardt \cite{metzvoll}
and subsequently M\"uller-Hartmann \cite{muha89}, Brandt and Mielsch \cite{bm}
and Jani\v{s} \cite{janis} it is known that a correlated lattice model can be
mapped onto an effective impurity system in the limit $d\rightarrow \infty$.
This is one consequence of the important aspect of this limit, namely that the
irreducible one-particle self energy is purely local \cite{metzvoll,muha89}
and a functional of the local propagator only \cite{bm,janis,Jarrell,gk}.
This property can be used to rewrite the lattice problem in such a way
that one is left with the solution of an effective
single-impurity Anderson model (SIAM), where the free bandstates are replaced
by an effective medium obtained from the full problem with the site under
consideration removed \cite{bm,janis,Jarrell,gk}. The one-particle Greens
function
or equivalently the one-particle self energy of the system are then given by
the corresponding quantities of the effective single-site problem. We shall
see later, that one can also calculate the two-particle correlation functions
of the lattice system with the help of those of the effective SIAM.
Note that this effective theory
preserves the dynamics introduced by the local correlations and thus
is still highly nontrivial since there does not exist a complete solution
for the SIAM. However, there exist at least different numerical exact
techniques
like quantum Monte Carlo and
controlled perturbational approximations to solve this local model
\cite{pjf,pcj93}. All these methods can then in turn be used to provide
a solution of correlated
lattice models {\em in the thermodynamical limit}. This approach
has become known as the {\em dynamical mean-field theory}. The name is based on
the observations that (i) the limit $d=\infty$
provides a canonical starting point for the construction of a thermodynamically
consistent mean-field theory \cite{itzykson} and (ii) in contrast to
a standard mean-field theory (like e.g.\ the one for the Heisenberg model)
one obtains a complex, frequency dependent function as molecular field
due to the dynamical nature of the local Coulomb repulsion.
Note that with the same arguments one also finds that the contribution to the
one-particle  self-energy
due to interactions like the spin exchange in the model (\ref{tjm}) is given by
the corresponding Hartree diagram only and thus is also purely local and
in addition static \cite{muha89}.
The latter statement means that for $d=\infty$ the $t$-$J$ model in the
paramagnetic phase (i.e. when $\langle S_z^i\rangle=0$) is identical to
the Hubbard model with $U=\infty$. Regarding the one-particle properties in
this regime we thus expect the well known features of the Hubbard model
\cite{pjf}. The situation of course changes as soon as one has a transition
into a magnetic state which will be discussed elsewhere \cite{to}.
\subsection{Susceptibility for the $t$-$J$ model}
For our purposes it is convenient
to represent the transverse spin susceptibility of the $t$-$J$ model as
\begin{equation}\label{hubabuba}
\chi_{\vec{q}}^{tJ}(i\nu_n)=\frac{1}{\beta^2}\suml_{\omega_n,\omega_m}
\chi_{\vec{q}}(i\omega_n,i\omega_m;i\nu_n)e^{i(\omega_n+\omega_m)0^+}\;\;,
\end{equation}
where $\chi_{\vec{q}}(i\omega_n,i\omega_m;i\nu_n)$ is the spatial Fourier
transform of the particle-hole propagator
\begin{eqnarray}\label{chiijdef}
\chi_{ij}(i\omega_n,i\omega_m;i\nu_{l}) =  \frac{1}{\beta}
\intl^{\beta}_{0} d\tau_{1} \intl^{\beta}_{0} d\tau_{2} \intl^{\beta}_{0}
d\tau_{3}
\intl^{\beta}_{0} d\tau_{4}\, e^{-i\omega_{m}(\tau_{1}-\tau_{2})}
e^{-i\omega_{n}(\tau_{3}-\tau_{4})} e^{-i\nu_{l}(\tau_{2}-\tau_{4})}
\nonumber \\
\langle T_{\tau} c_{i \uparrow }(\tau_{4})
 c^{+}_{i \downarrow}(\tau_{3}) c_{j \downarrow}(\tau_{2}) c^{+}_{j\uparrow }(
\tau_{1})\rangle_{tJ}\;\;.
\end{eqnarray}
In equations (\ref{hubabuba}) and (\ref{chiijdef}) $i\omega_n$ and $i\omega_m$
denote Fermi Matsubara frequencies
and $i\nu_n$ a Bose Matusbara frequency.
Quite generally, by introducing the irreducible two-particle self energy
$\Gamma_{ij}^{\uparrow\downarrow}(i\omega_n,i\omega_m;i\nu_{l})$, the
particle-hole propagator
(\ref{chiijdef}) can formally be written as
\begin{equation}
\chi_{ij}(i\omega_n,i\omega_m;i\nu_{l}) =
\beta\chi^{(0)}_{ij}(\begin{array}[t]{l}\D\! i\omega_n;i\nu_{l})
\delta_{n,m} \\[5mm]
\D+ \frac{1}{\beta}
\sum_{lk,i\omega_{p}}\chi^{(0)}_{il}(i\omega_n;i\nu_{l})
 \Gamma_{lk}^{\uparrow\downarrow}(i\omega_n,i\omega_p;i\nu_{l})
\chi_{kj}(i\omega_p,i\omega_m;i\nu_{l})\;\;.\end{array}
\label{vertex}
\end{equation}
Here, $\chi^{(0)}_{ij}(i\omega_n;i\nu_n)=-
G_{ij}(i\omega_n)G_{ji}(i\omega_n+i\nu_n)$ represents
the unperturbed part of the particle-hole propagator and
$G_{ij}(i\omega_n)$ the full one-particle Greens function of the system.

Using standard techniques of
field theory \cite{bm}, one can express the irreducible particle-hole self
energy as
functional derivative of the one-particle self energy with respect to the
one-particle propagator. In combination with the observation, that within the
dynamical mean-field theory (DMFT) (i) the one-particle self energy is purely
local and (ii) the exchange term $\Jast$ enters the one-particle self energy
only on the Hartree level it follows that the two-particle self energy
aquires the particularly simple form
\begin{equation}\label{twopse}
\Gamma_{lk}^{\uparrow\downarrow}(i\omega_n,i\omega_p;i\nu_{l})
=
-\frac{\Jast}{\cal Z}\delta_{|i-j|,{\rm n.N.}}+
\Gamma^{\uparrow\downarrow}(i\omega_n,i\omega_p;i\nu_{l})\;\;.
\end{equation}
The non-trivial second term is the irreducible particle-hole self energy for
$\Jast=0$, i.e.\ for the $U=\infty$-Hubbard model.
Note that within the DMFT this quantity is also purely local \cite{bm}!

Inserting the result (\ref{twopse}) into the expression (\ref{vertex}) and
transforming into $\vec{q}$-space, we obtain as transverse
magnetic susceptibility of the $t$-$J$ model in the DMFT
\begin{equation}\label{chitJ}
\chi_{\vec{q}}(i\omega_n,i\omega_m;i\nu_n) =
\beta\chi^{(0)}_{\vec{q}}\begin{array}[t]{l}\D(i\omega_n;i\nu_n)\delta_{nm}
+
J_{\vec{q}}\chi^{(0)}_{\vec{q}}(i\omega_n;i\nu_n)
\frac{1}{\beta}\suml_p\chi_{\vec{q}}(i\omega_p,i\omega_m;i\nu_n)\\[5mm]
\D+\frac{1}{\beta}\suml_p\chi^{(0)}_{\vec{q}}(i\omega_n;i\nu_n)
\Gamma^{\uparrow\downarrow}(i\omega_n,i\omega_p;i\nu_n)
\chi_{\vec{q}}(i\omega_p,i\omega_m;i\nu_n)\;\;.\end{array}
\end{equation}
In equation (\ref{chitJ}) $J_{\vec{q}}$ denotes the Fourier transform of
$\displaystyle-\frac{\Jast}{\cal Z}\delta_{|i-j|,{\rm n.N.}}$. For the case of
a
simple hyper-cubic lattice
one e.g.\ obtains $\displaystyle
J_{\vec{q}}=-\frac{\Jast}{d}\suml_{l=1}^d\cos(q_l\cdot a)$.

The susceptibility (\ref{chitJ}) contains as one contribution the
susceptibility
of the Hubbard model in the limit $U=\infty$ given by \cite{Jarrell}
\begin{equation}\label{chiHM}
\chi_{\vec{q}}^{HM}(i\omega_n,i\omega_m;i\nu_n) =
\begin{array}[t]{l}\D
\beta\chi^{(0)}_{\vec{q}}(i\omega_n;i\nu_n)\delta_{nm}\\[5mm]
\D+\frac{1}{\beta}\suml_p\chi^{(0)}_{\vec{q}}(i\omega_n;i\nu_n)
\Gamma^{\uparrow\downarrow}(i\omega_n,i\omega_p;i\nu_n)
\chi_{\vec{q}}^{HM}(i\omega_p,i\omega_m;i\nu_n)\;\;.\end{array}
\end{equation}
It is now straightforward to show that with the help of expression
(\ref{chiHM})  equation
(\ref{chitJ}) can be rewritten as
\begin{equation}\label{chitJ2}
\chi_{\vec{q}}(i\omega_n,i\omega_m;i\nu_n) = \begin{array}[t]{l}\D
\chi^{HM}_{\vec{q}}(i\omega_n,i\omega_m;i\nu_n)\\[5mm]
\D+J_{\vec{q}}\frac{1}{\beta}\suml_l\chi^{HM}_{\vec{q}}(i\omega_n,i\omega_l;i\nu_n)
\frac{1}{\beta}\suml_p\chi_{\vec{q}}(i\omega_p,i\omega_m;i\nu_n)\;\;.\end{array}
\end{equation}
Performing the sums on $n$ and $m$ in equation (\ref{chitJ2}) finally
leads to the appealing result
\begin{equation}\label{chitJ_final}
\begin{array}{l}\D
\chi_{\vec{q}}^{tJ}(i\nu_n)=\chi_{\vec{q}}^{HM}(i\nu_n)
+J_{\vec{q}}\chi_{\vec{q}}^{HM}(i\nu_n)\chi_{\vec{q}}^{tJ}(i\nu_n)\\[5mm]
\chi_{\vec{q}}^{tJ}(i\nu_n)=\chi_{\vec{q}}^{HM}(i\nu_n)\left[
1-J_{\vec{q}}\chi_{\vec{q}}^{HM}(i\nu_n)\right]^{-1}
\end{array}
\end{equation}
as expression for the magnetic susceptibility of the $t$-$J$ model in the
DMFT.
Thus the major ingredient in the susceptibility of the $t$-$J$ model is
the corresponding quantity of the HM for $U=\infty$. One should also note that
the expression (\ref{chitJ_final}) is very similar to the standard RPA
result
\begin{equation}\label{chiU0}
\chi_{\vec{q}}(i\nu_n;U=0)=\chi_{\vec{q}}(i\nu_n;U=0,J=0)\left[
1-J_{\vec{q}}\chi_{\vec{q}}(i\nu_n;U=0,J=0)\right]^{-1}
\end{equation}
for the corresponding noninteracting system. Thus, as far as the DMFT for the
$t$-$J$ model is concerened, the susceptibility is formally obtained by simply
replacing
$\chi_{\vec{q}}(i\nu_n;U=0,J=0)$ by $\chi_{\vec{q}}(i\nu_n;U=\infty,J=0)$ in
the
RPA-formulas. Let us emphasize that this correspondence holds only on a formal
level: The physical situation described by (\ref{chitJ_final}) is of course
fundamentally different from the one modeled by (\ref{chiU0})!
\subsection{The spin susceptibility of the Hubbard model}
As already mentioned,
the dynamic spin susceptibility of the Hubbard model in real space is
within the DMFT
given by \cite{Jarrell}
\begin{equation}\label{chiijHM}
\chi_{ij}^{HM}(i\omega_n,i\omega_m;i\nu_n) =
\begin{array}[t]{l}\D
\beta\chi^{(0)}_{ij}(i\omega_n;i\nu_n)\delta_{nm}\\[5mm]
\D+\frac{1}{\beta}\suml_{l,i\omega_p}\chi^{(0)}_{il}(i\omega_n;i\nu_n)
\Gamma^{\uparrow\downarrow}(i\omega_n,i\omega_p;i\nu_n)
\chi_{lj}^{HM}(i\omega_p,i\omega_m;i\nu_n)\;\;.\end{array}
\end{equation}
Equation (\ref{chiijHM}) obviously also holds for the local susceptibility,
i.e.
\begin{equation}
\chi_{loc}(i\omega_n,i\omega_m;i\nu_{l}) =
\chi_{loc}^{(0)}(i\omega_n;i\nu_{l})\,[\,\beta\delta_{n,m}
+ \frac{1}{\beta} \sum_{i\omega_{p}}
\Gamma^{\uparrow\downarrow}(i\omega_n,i\omega_p;i\nu_{l})
\chi_{loc}(i\omega_p,i\omega_m;i\nu_{l})\,]
\label{localsus}
\end{equation}
with the same $\Gamma^{\uparrow\downarrow}(i\omega_n,i\omega_p;i\nu_{l})$ as in
equation (\ref{chiijHM}).
Combining equations (\ref{chiHM}) and (\ref{localsus}), the susceptibility can
be
expressed
by the local susceptibility through a matrix equation
($[\mat{A}_{\vec{q},l}]_{nm}
=A_{\vec{q}}(i\omega_n,i\omega_m;i\nu_l)$)
\begin{equation}\label{chiqHM}
\begin{array}{l}
\D\mat\chi_{\vec{q},l}=\left[\mat1-\frac{1}{\beta}
\mat\chi_{loc,l}\cdot\mat\Gamma_{\vec{q},l}^{eff}\right]^{-1}
\mat\chi_{loc,l}\;\;,\\[5mm]
\D\mat\Gamma_{\vec{q},l}^{eff}=-({\mat\chi_{\vec{q},l}^{(0)}}^{-1} -
{\mat\chi_{loc,l}^{(0)}}^{-1})\;\;.
\end{array}
\end{equation}
With the definition $\left[\vec{\Lambda}_l\right]_m =
\Lambda_l(i\omega_m)=\frac{1}{\beta}\suml_{n}
\chi_{loc}(i\omega_n,i\omega_m;i\nu_l)$ and the symmetry relation
$\chi_{loc}(i\omega_n,i\omega_m;i\nu_l)=\chi_{loc}(i\omega_m,i\omega_n;-i\nu_l)$
following from the definition (\ref{chiijdef}) we can formally perform the
frequency sums in (\ref{chiqHM}) to obtain
\begin{equation}\label{chiHMfinal}
\chi_{\vec{q}}^{HM}(i\nu_l)=\chi_{loc}^{HM}(i\nu_l)
+\frac{1}{\beta}\vec{\Lambda}_l^T\cdot\mat\Gamma_{\vec{q},l}^{eff}
\frac{\mat1}{\mat1-\frac{1}{\beta}\mat\chi_{loc,l}\cdot\mat\Gamma_{\vec{q},l}^{eff}}\cdot
\vec{\Lambda}_{-l}
\end{equation}
as the final result for the magnetic susceptibility of the Hubbard model in
the framework of the dynamical molecular field theory.

It is important to note that until now no explicit reference to the value
of $U$ has been made, i.e.\ equation (\ref{chiHMfinal}) is valid for all $U$.
The form (\ref{chiHMfinal}) for the susceptibility of the HM
is especially convenient for computational reasons, because the outer sums on
Matsubara frequencies
have been performed exactly. These can pose numerical problems because
\mbox{$[\mat\chi_{\vec{q},l}]_{nm}$} decays at most like $1/(nm)$ for
large $n,m$ and one has to care for the correct time ordering in the final sums
(cf.\ equations (\ref{chitJ}) and (\ref{chiijHM})). Whereas for the inner sums
the products occuring there lead to an asymptotic behaviour like at least
$\sim 1/n^2$ and thus a well defined sum.
\subsection{The local spin susceptibility}
The only unkown quantity in equation (\ref{chiHMfinal}) is the local
susceptibility $\chi_{loc}(i\omega_n,i\omega_m;i\nu_n)$ defined by
\begin{eqnarray}\label{blablabla}
\chi_{loc} (i\omega_{n}, i\omega_{m}; i\nu_{n}) =  \frac{1}{\beta}
\intl^{\beta}_{0} d\tau_{1} \intl^{\beta}_{0} d\tau_{2} \intl^{\beta}_{0}
d\tau_{3}
\intl^{\beta}_{0} d\tau_{4}\, e^{-i\omega_{m}(\tau_{1}-\tau_{2})}
e^{-i\omega_{n}(\tau_{3}-\tau_{4})} e^{-i\nu_{n}(\tau_{2}-\tau_{4})}
\nonumber \\
< T_{\tau} c_{i \uparrow }(\tau_{4})
 c^{+}_{i \downarrow}(\tau_{3}) c_{i \downarrow}(\tau_{2}) c^{+}_{i\uparrow
}(\tau_{1}) >
\label{lds}
\end{eqnarray}
Within the DMFT, this function is obtained from the corresponding quantity
of an effective SIAM with the band electrons replaced by the effective
medium of the DMFT.

For finite U, the most successful way to solve the effective single-site
problem and calculate functions like (\ref{blablabla}) is by Quantum
Monte Carlo techniques \cite{Jarrell}. However, since we are interested in
the limit $U=\infty$ in the current context, this technique is not available.
On the other hand, for $U=\infty$ the time-ordered perturbation theory
\cite{Keiter} provides a natural and easy access to local quantities.
In this method one expresses all local quantities through the resolvents
$P_{0(1\sigma)}(z)$ of the unoccupied (occupied) ionic states.
Of course, this theory cannot be solved exactly, so further approximations
have to be introduced. Here, we shall use the so-called non-crossing
approximation
(NCA) \cite{Keiter,coxbickerswilkins} to calculate these resolvents and
express further local correlation functions of interest.
In previous publications
we have already shown that the NCA
provides a reliable approximation scheme
to calculate such local quantities \cite{pjf,Jarrell,pcj93}.
Applying the standard diagrammatic rules of this perturbational technique
\cite{Keiter} in conjunction with the NCA we obtain
\begin{eqnarray}\label{chilocNCA}
\chi_{loc} (i\omega_{n}, i\omega_{m};i\nu_l)  =  -
\frac{1}{Z_{loc}}\oint\limits_{{\cal C}} \frac{dz}{2\pi i}  e^{-\beta z}
P_{1}(z) P_{1}(z-i\nu_l) P_{0}(z-i\omega_{n}) P_{0}(z-i\omega_{m}-i\nu_l)
\end{eqnarray}
for the local susceptibility.
In equation (\ref{chilocNCA}), $Z_{loc}=\sum\limits_M\oint\limits_{{\cal C}}
\frac{dz}{2\pi i}  e^{-\beta z}P_M(z)$ denotes the local contribution to
the partition function and the contour ${\cal C}$ surrounds all singularities
of the integrands counterclockwise.

\section{Results}
\subsection{General remarks}
The expressions (\ref{chiHMfinal}) and (\ref{chilocNCA}) in principle still
allow for the calculation of the dynamical
susceptibility. Unfortunately, the derivation of equation (\ref{chiHMfinal})
utilizes the representation of all quantities in Matsubara-space, i.e.\ one
would
be left with the awkward task to analytically continue the results to the
real axis. This nontrivial problem is left for a future publication
\cite{jpfuture}.
In this contribution we will concentrate on the static susceptibility, i.e.\ we
set $i\nu_l=0$.

Before we turn to the actual results for the $U=\infty$-Hubbard and $t$-$J$
model
let us first briefly discuss the special limit $\langle n\rangle=1$. In
this case the model (\ref{tjm}) becomes equivalent to the Heisenberg model and
it is a straightforward task to calculate the molecular field expression for
the static susceptibility, which reads
\begin{equation}\label{chiq_half}
\chi_{\vec{q}}^{n=1}=\frac{\frac{\beta}{2}}{1-J_{\vec{q}}
\frac{\beta}{2}}\;\;.
\end{equation}
Comparing this expression with the result for the $t$-$J$ model in equation
(\ref{chitJ_final}), one sees that obviously
$\chi_{\vec{q}}^{HM}\to\frac{\beta}{2}$ for $\langle n\rangle\to1$. On the
other hand, $\beta/2$
is also exactly the value we expect for the {\em local}\/ susceptibility in
this
limit, i.e.\ $\chi_{loc}^{HM}\to\frac{\beta}{2}$ for $\langle n\rangle\to1$.
 From this it at once follows that the second part in equation
(\ref{chiHMfinal})
will become negligible for $\langle n \rangle$ close to half filling. On the
one hand this offers a rather sensible test for the numerics involved in
calculating
the susceptibility for the HM. In addition it provides an interesting
approximate ansatz for the susceptibility of the $t$-$J$ model
by setting $\chi_{\vec{q}}^{HM}\approx\chi_{loc}^{HM}$ in this limit.
Note that this also allows for a simple approximate calculation of dynamics
since $\chi_{loc}^{HM}(\omega)$ is much easier to obtain than
$\chi_{\vec{q}}^{HM}(\omega)$ given by (\ref{chiHMfinal}). The latter
observation
is especially interesting in the light of recent studies by Scalapino et al.\
who
analyzed the dynamical susceptibility for the twodimensional $t$-$J$ model
obtained from exact diagonalization and found that it was rather well described
by a
form like (\ref{chitJ_final}) with $\chi_{\vec{q}}^{HM}(\omega)$ replaced
by some local quantity \cite{scalapino}.

\subsection{The Hubbard model}
Let us start by discussing the Lindhardt function
\begin{equation}\label{lindhardt}
\chi^{(0)}_{\vec{q}}=-\frac{1}{N\beta}\suml_{\omega_n,\vec{k}}
G_{\vec{k}+\vec{q}}(i\omega_n)G_{\vec{k}}(i\omega_n)\;\;.
\end{equation}
While the whole derivation was completely independent of the actual lattice
structure, we now have to specify the meaning of the $\vec{k}$-sum. We here
choose a simple cubic lattice in $d$ dimensions, i.e.\ the coordination number
is ${\cal Z}=2d$,  and take the limit $d\to\infty$
to use the simplifications arising in this limit \cite{muha89}. With
$\tast=1$ as the unit of energy, one then obtains
for the single-particle DOS the well-known Gaussian form
$\rho_0(\epsilon)=\exp(-\epsilon^2)/\sqrt{\pi}$ \cite{muha89} and one can
also evaluate the $\vec{k}$-sum in equation (\ref{lindhardt}) analytically
\cite{muha89,bm} to yield
{\small
\begin{equation}\label{lindhardt2}
\chi^{(0)}_{\vec{q}}=
\frac{1}{\beta}\suml_{\omega_n}\intl_{-\infty}^\infty
d\epsilon d\epsilon'
\frac{\rho_0(\epsilon)\rho_0(\epsilon')}{\left(i\omega_n+\mu-
\Sigma(i\omega_n)-\epsilon\right)\cdot\left(i\omega_n+\mu-\Sigma(i\omega_n)-
\epsilon\cdot\eta_{\vec{q}}-\epsilon'\cdot\sqrt{1-\eta_{\vec{q}}^2}\right)}\;\;.
\end{equation}}
In relation (\ref{lindhardt2}), $\eta_{\vec{q}}=\suml_{l=1}^d\cos(q_l\cdot
a)/d$ and
$\Sigma(z)$ is the one-particle self energy of the HM for a given $U\ge0$.
Note that the external wave-vector
$\vec{q}$ only enters via the function $\eta_{\vec{q}}$ which basically
describes surfaces of constant energy in the simple cubic Brillouin zone.
For presentational reasons, we shall choose the special vector
$\vec{q}=q(1,1,1,1,\ldots)$ and use the number $q$ with $0\le q\le\pi$ as label
rather than
$-1\le\eta_{\vec{q}}\le 1$.

\begin{figure}[htb]
\centerline{\psfig{figure=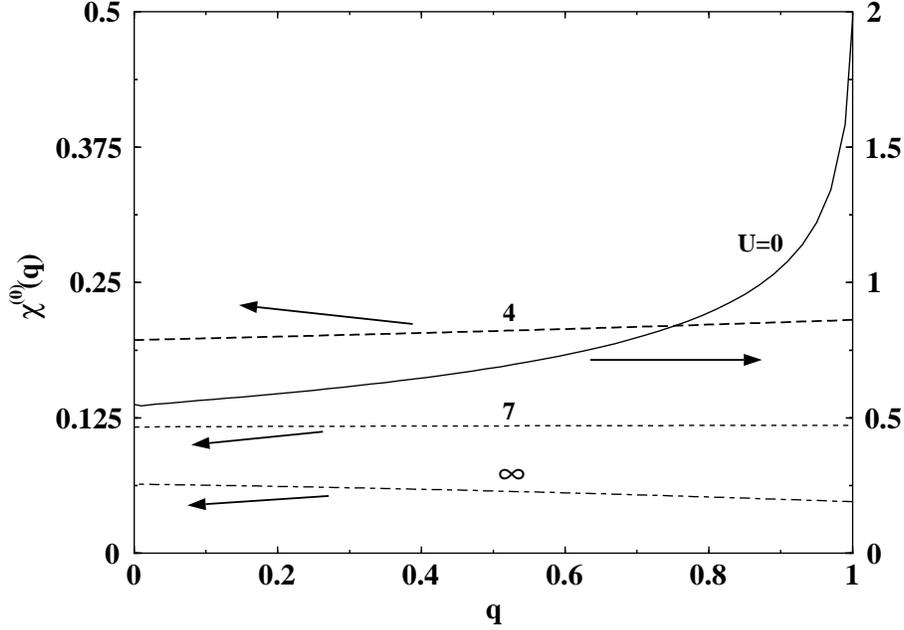,width=12cm}\ }
\caption[]{Lindhardt function for $U=0$ ,$4$, $7$ and $U=\infty$}
\label{chiq0}
\end{figure}
The Lindhardt function for the HM for four different values of $U=0$, $4$, $7$
and $U=\infty$
at a filling $\langle n\rangle=0.95$ and for a low temperature $T=1/30$ is
shown in Fig.~\ref{chiq0}. Note the different scales for $U=0$
(right scale) and $U=4$ ,$7$ and $U=\infty$ (left scale)! Without looking at
the details
it is thus clear, that the correlations induced by $U$ strongly suppress
this quantity. In addition one can observe a dramatic change in the
$q$-dependence
with increasing $U$. While for $U=0$ one has a strong peak at $q=\pi$ due to
the nesting property of the simple-cubic Fermi surface close to half filling
this feature is strongly suppressed by the damping introduced by the
correlations for $U=4$ ,$7$ and $U=\infty$. In addition there occurs a
cross-over from the maximum in
$\chi^{(0)}_{\vec{q}}$ being at $q=\pi$ for small $U$ to $q=0$ for $U=\infty$.
Note also that in contrast to $U=0$ the total $q$-dependence is rather weak in
the other cases.

 From the previous observation one may deduce two things: First,
since for $U=\infty$ there is no net magnetic exchange between neighbouring
sites, we expect from the flatness of $\chi^{(0)}_{\vec{q}}$ that also
$\chi^{HM}_{\vec{q}}$ will be relatively flat as a function of $\vec{q}$. In
addition, the fact that $\chi^{(0)}_{\vec{q}}$ is maximal at $q=0$ suggests
that
$\chi^{HM}_{\vec{q}}$ for $U=\infty$ will be enhanced at $q=0$ rather than
at $q\approx\pi$ as expected and observed for $U<\infty$ \cite{Jarrell}.

\begin{figure}[htb]
\centerline{\psfig{figure=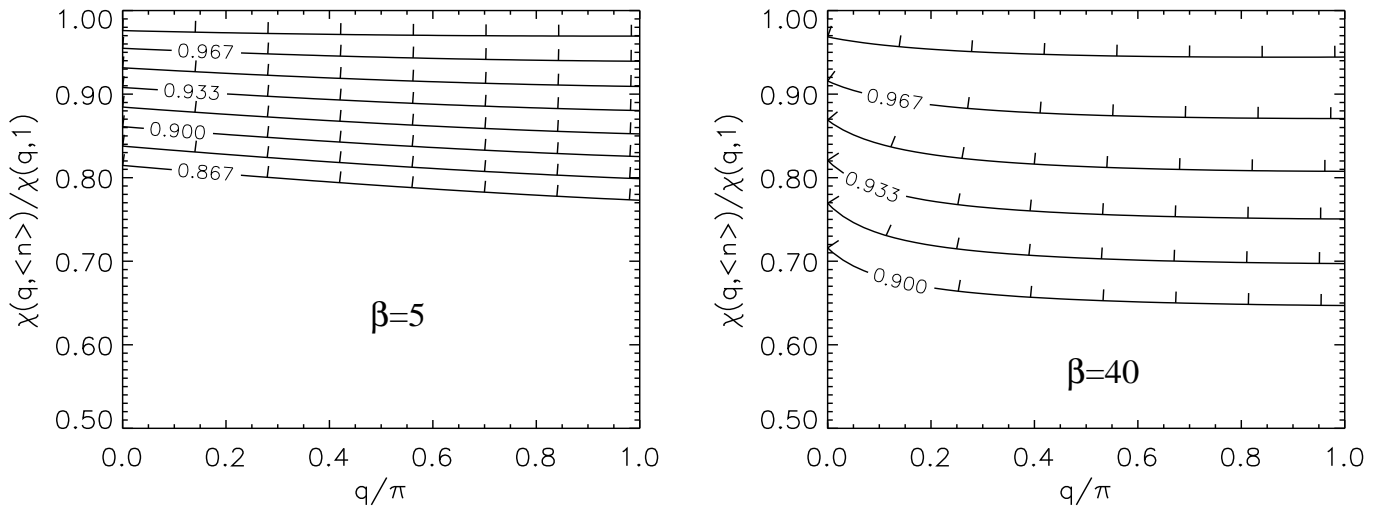,width=18cm}\ }
\caption[]{Susceptibility of the HM at $U=\infty$ as function of $\vec{q}$ and
filling for two different temperatures $T=1/5$ and $T=1/40$.}
\label{chiqofnT}
\end{figure}
This behaviour can indeed be seen in Fig.~\ref{chiqofnT}, where we have
plotted $\chi^{HM}_{\vec{q}}$ for two different temperatures as function
of $q$ and doping $\delta=1-\langle n\rangle$. The susceptibility was
normalized to its value at
$\delta=0$, i.e.\ to $\chi_{\vec{q}}^{HM}(\delta=0)=\beta/2$.
Note that we always find $\chi^{HM}_{\vec{q}}(\delta>0) <
\chi_{\vec{q}}^{HM}(\delta=0)$.
 From the form (\ref{chitJ_final}) for the susceptibility of the $t$-$J$ model
it then at once follows that also $\chi^{tJ}_{\vec{q}}(\delta>0) <
\chi^{tJ}_{\vec{q}}(\delta=0)$ for all values of $\Jast$ and $\vec{q}$. This
should be compared with results from high-temperature expansions for $d=2$
\cite{puttica} which suggest a pronounced maximum in the uniform susceptibility
around $\delta=15$\% produced by spin fluctuations not included in the current
mean-field treatment.

\begin{figure}[htb]
\centerline{\psfig{figure=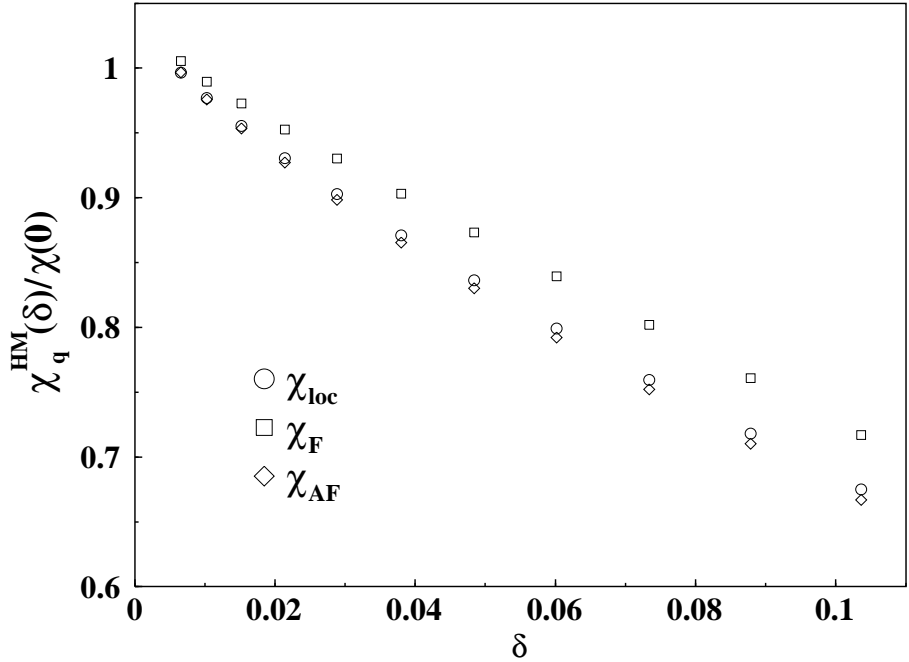,width=12cm}\ }
\caption[]{Susceptibility for $U=\infty$ and $\beta=30$ as function of
$\delta$.}
\label{chiofd}
\end{figure}

Another interesting feature in Fig.~\ref{chiqofnT} is that
in all cases the variation with $q$ is comparatively weak,
becoming somewhat stronger for lower temperatures and with increasing doping
$\delta$. We also observe a slight maximum at $q=0$ that
becomes more pronounced for lower temperatures but interestingly weakens
with decreasing doping for $T$ fixed. This observation is substantiated by a
look at the doping dependence of $\chi_{\vec{q}}^{HM}$ in Fig.~\ref{chiofd} for
the
local (circles), ferromagnetic $q=0$ (squares) and antiferromagnetic
$q=\pi$ (diamonds)
susceptibility for an inverse temperature $\beta=30$. It is interesting to note
that the antiferromagnetic
susceptibility of the HM at $U=\infty$ is always very close to the local one,
which can be understood by the fact that due to the mapping of the HM onto an
equivalent impurity model the local susceptibility already contains most of
the (nearest-neighbour) antiferromagnetic correlations. Since for $U=\infty$
there is no additional net magnetic exchange the nonlocal corrections only
give a small renormalization. In contrast to this the renormalizations for
the ferromagnetic susceptibility are comparatively strong and definitely tend
to enhance this quantity above both the local and antiferromagnetic
susceptibility.
These results have to be interpreted
in the light of Nagaoka's theorem \cite{nagaoka}, where in the presence of
{\em one hole}\/ a ferromagnetic state for the background
is favoured from a minimization of the hopping energy in the correlated
system but not as a result of a direct magnetic coupling. Obviously, our
results
suggest that sizeable ferromagnetic correlations still exist for a finite
number of holes. However, so far we do not find any hint towards a
ferromagnetic
instability at low temperatures close to half filling. This is consistent
with the conjecture that for bipartite lattices -- like the simple hyper-cubic
lattice studied here --
the critical hole density for the Nagaoka state should be $\delta_c=0$
\cite{mueha95}.
\subsection{Results for the $t$-$J$ model}
\begin{figure}[htb]
\centerline{\psfig{figure=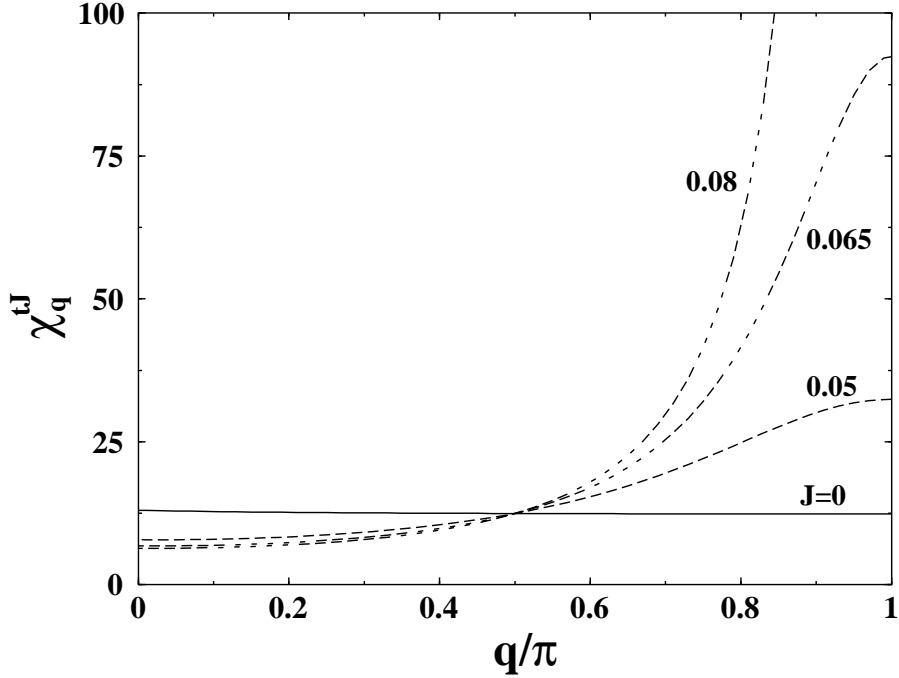,width=12cm}\ }
\caption[]{Susceptibility of the $t$-$J$ model as function of $\vec{q}$ for
various
values of $J$ at a doping $\delta=5$\% and $\beta=30$.}
\label{chiqtJ}
\end{figure}
Inserting the results for the susceptibility of the HM at $U=\infty$ into
equation (\ref{chitJ_final}) we obtain the susceptibility for the $t$-$J$
model as function of $q$ and $\Jast$ as shown in Fig.~\ref{chiqtJ} for $\langle
n\rangle=0.95$
and $\beta=30$. The explicit exchange now obviously favours the
antiferromagnetic point
$q=\pi$ and eventually leads to an antiferromagnetically ordered state for
$\Jast
>\Jast_c\approx0.085$
for this particular parameter set.

\begin{figure}[htb]
\centerline{\psfig{figure=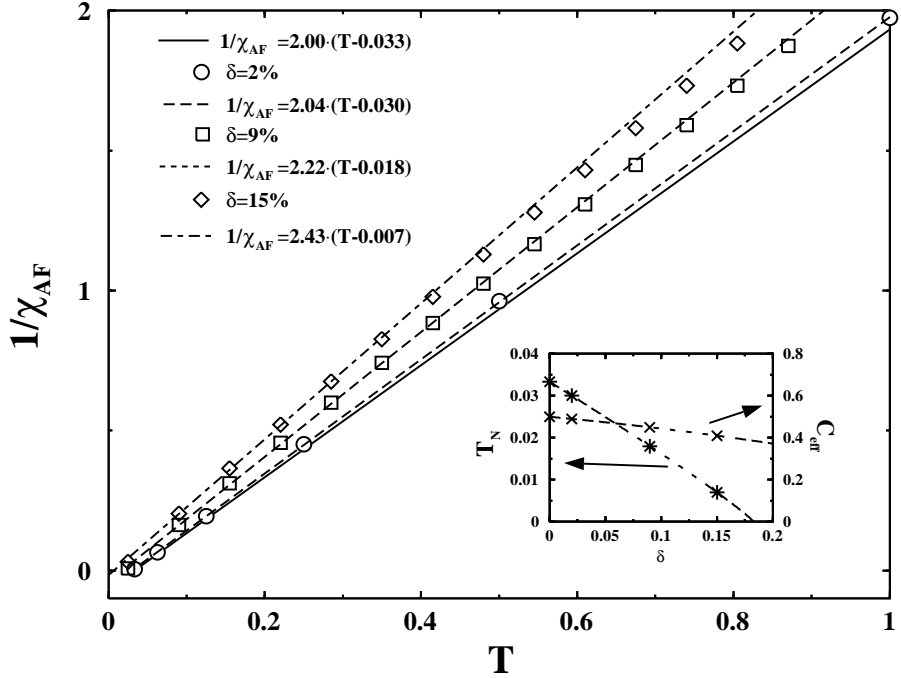,width=12cm}\ }
\caption[]{Inverse susceptibility of the $t$-$J$ model as function of $T$ for
$\Jast=0.067$
and three dopings $\delta=2$\%, $\delta=9$\% and $\delta=15$\%. Close
to the phase transition one observes $\chi_{AF}^{-1}(T)= (T-T_N)/C_{eff}$ as
expected
for a mean-field theory. Note that for $\delta\to0$ the linear behaviour
is observed up to $T=1\tast$. The full line represents half filling, where
$\chi_{AF}^{-1}=2\cdot(T-\Jast/2)$. The inset shows the dependence of the
N\'eel
temperature $T_N$ and effective Curie constant
$C_{eff}$ on $\delta$.}
\label{chiofTJ}
\end{figure}
The temperature dependence of $1/\chi^{tJ}_{AF}$ for a specific value of
$\Jast=0.067$ and three dopings $\delta=2$\%, $\delta=9$\% and
$\delta=15$\% is collected in Fig.~\ref{chiofTJ}. The
full curve marks for comparison the case $\delta=0$, where one
has exactly $1/\chi^{tJ}_{AF}=2\cdot(T-\Jast/2)$. As
expected for a mean-field theory, close to the antiferromagnetic transition one
finds a behaviour
$1/\chi^{tJ}_{AF}=  (T-T_N)/C_{eff}$ in all cases with decreasing N\'eel
temperature $T_N$ and decreasing effective Curie constant $C_{eff}$ for
increasing $\delta$ (see e.g.\ inset to Fig.~\ref{chiofTJ}). It is quite
noteworthy that close to half filling
(i.e.\ for $\delta=2$\%) this linearity extends up to rather high temperatures.
However, with increasing doping one eventually finds appreciable deviations
from this
linearity for temperatures well above $T_N$. Both $T_N$
and $C_{eff}$ vary roughly linear up to $15$\% doping.
We would also like to point out that up to a doping of $\delta=15$\% we do not
observe any tendency towards incommensurate order.

\begin{figure}[htb]
\centerline{\psfig{figure=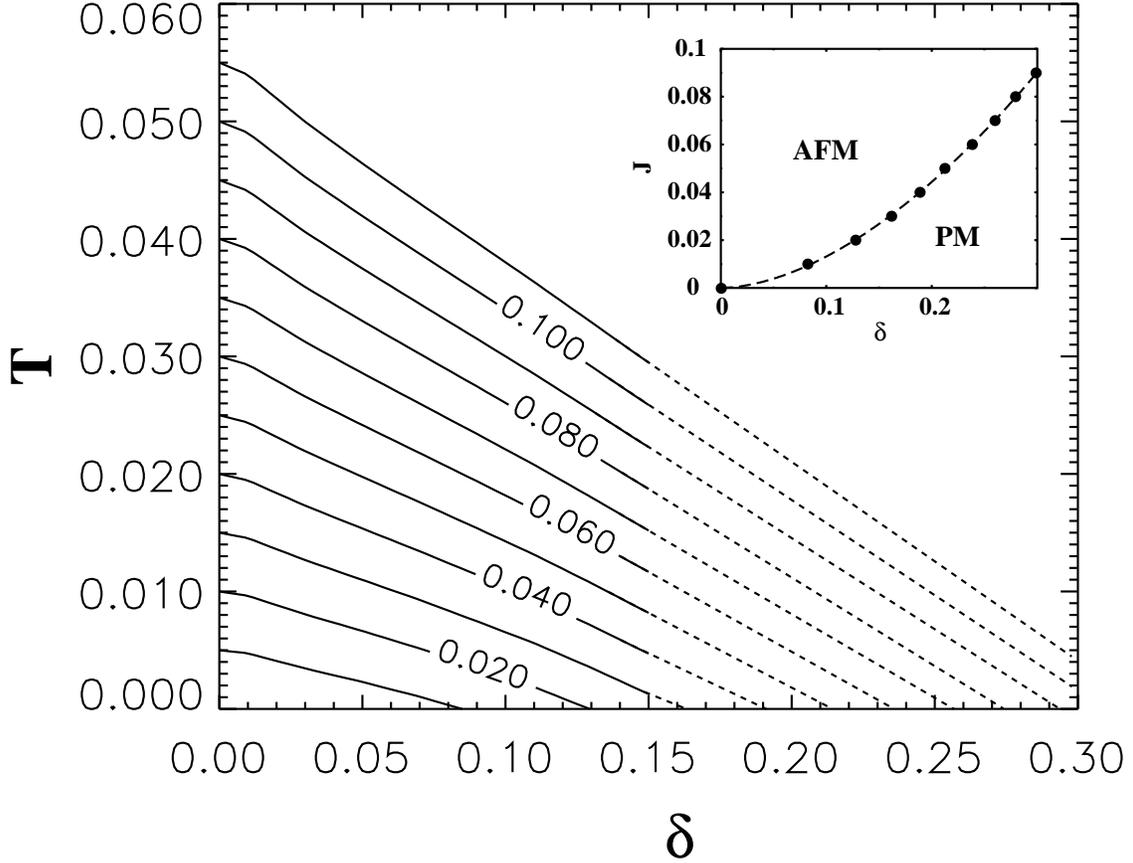,width=15cm}\ }
\caption[]{Phase diagram $T_N(\delta,J)$ for the $t$-$J$ model.
The dashed lines represent (linear) extrapolations of the phase boundaries
to $T=0$. The corresponding values of $J_c(\delta)$ behave like
$J_c(\delta)\sim\protect\delta^2$, as shown in the inset.}
\label{phaseb30}
\end{figure}
With the method outlined above we are now able to calculate the phase diagram
$T_N(\delta,\Jast)$ for the $t$-$J$ model. The results for
dopings $\delta\le15$\% and $\Jast<0.12$ are shown in Fig.~\ref{phaseb30}.
One observes the expected increase in the N\'eel temperature
$T_N$ with increasing $\Jast$ and a -- for larger $\delta$ roughly linear
-- decrease as function of $\delta$.
We may use this approximate
linearity of $T_N(\delta)$ to extrapolate the curves $T_N(\delta)$ for a given
$\Jast$ to $T=0$.
This procedure allows us to obtain an extrapolation for the phase diagram
$\Jast_c(\delta)$
of the $t$-$J$ model at $T=0$. The result is shown in the inset to
Fig.~\ref{phaseb30}. We find that $\Jast_c(\delta)$ behaves rather accurately
like $\Jast_c\sim\delta^2$.
The phase diagram in Fig.~\ref{phaseb30} should be compared to the
DMFT results for the Hubbard model in the strong coupling limit
\cite{jar_free_94}. In reference \cite{jar_free_94} the authors calculate
$T_N(\delta,U)$ up to
$U=7\tast$, which would correspond to $\Jast\approx0.14$ for the $t$-$J$ model.
They also observe an almost linear dependence of $T_N$ on the doping $\delta$
for large values of $U$. However, although the value of $T_N$ for
$\delta\to0$ and the observed linearity agrees quite well with our results, the
depression of $T_N$ as
function of $\delta$ for the Hubbard model at $U=7\tast$ is much faster than in
our Fig.~\ref{phaseb30}.
In addition one encounters a transition into an incommensurate state for
$\delta\agt12$\%
in the Hubbard model.
Currently it is not clear whether these deviations -- especially the lack
of an incommensurate magnetic order for large doping -- between the results
for the large-$U$ Hubbard model and the $t$-$J$ model are real or due to the
additional
approximations introduced by using the NCA to solve the effective impurity
problem.
One should keep in mind, though, that for finite $U$ respectively $\Jast$ the
Hubbard model and the $t$-$J$ model are expected to show different physical
behaviour: The mapping of the Hubbard model to an effective model with
magnetic exchange generates in addition to the exchange term included in
the $t$-$J$ model also more complicated couplings, like for instance a
three-site term which is also
of the order $\Jast$ \cite{gross87} and may give rise to quite important
corrections in physical quantities \cite{ogata2}.

Finally we should like to use the observation that
\begin{figure}[htb]
\centerline{\psfig{figure=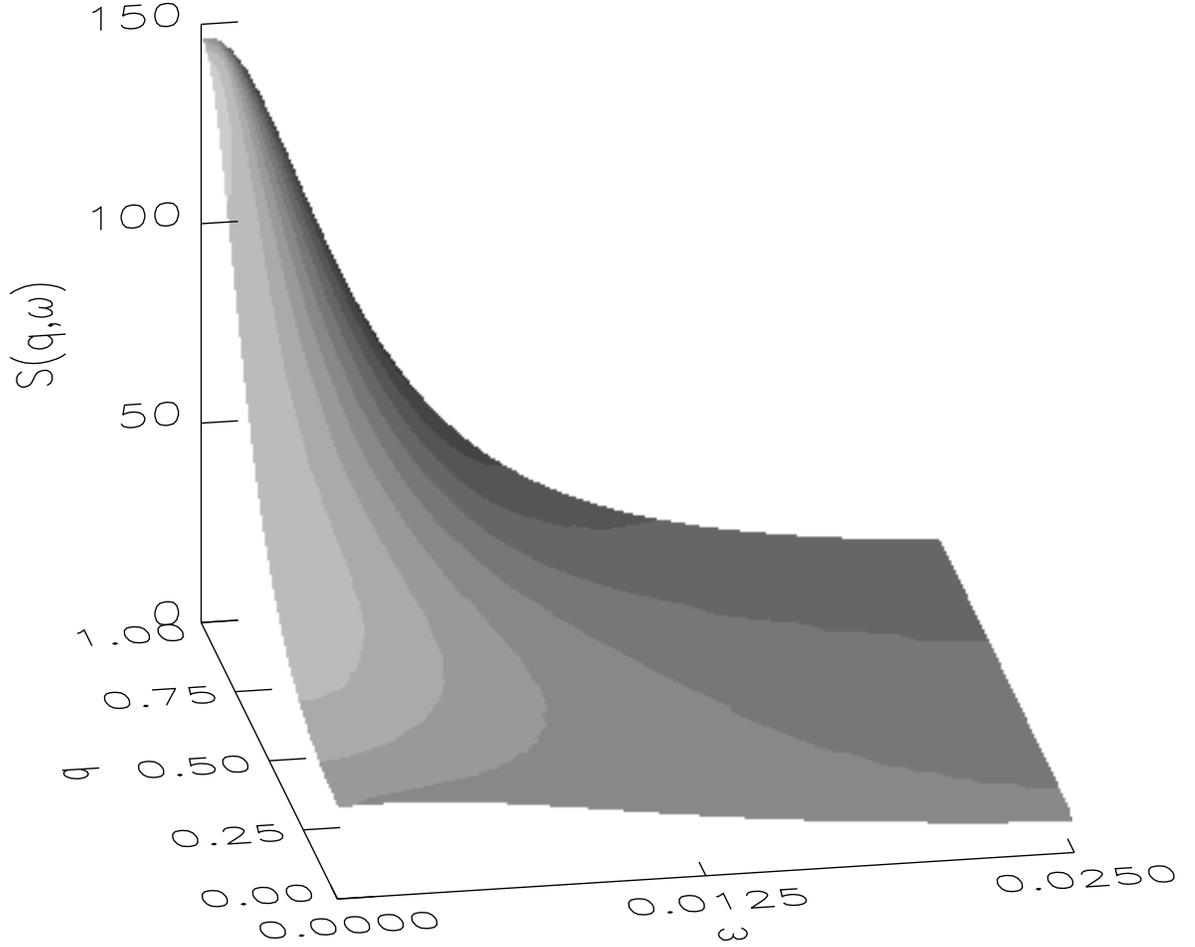,width=18cm}\ }
\caption[]{Approximate result for $S_{\vec{q}}^{tJ}(\omega)$ for $\delta=5$\%,
$T=1/30$ and $J=0.035$.}
\label{chiofwq}
\end{figure}
close to half filling the susceptibility for the HM is relatively flat with
respect to $\vec{q}$ and obtain an approximation for the dynamical
spin structure factor
$S(\vec{q},\omega)=\Im m\chi_{\vec{q}}(\omega)/(1-e^{-\beta\omega})$ by
assuming
$\chi^{HM}_{\vec{q}}(\omega)\approx\chi^{HM}_{loc}(\omega)$ in equation
(\ref{chitJ_final}).
This approximation avoids the cumbersome calculation of the $\vec{q}$-dependent
susceptibility for finite frequencies.
As an example the result for $\Jast=0.035$, $T=1/30$ and $\delta=5$\% is shown
in
Fig.~\ref{chiofwq}. As expected, the maximum in $S^{tJ}_{\vec{q}}(\omega)$
is found at $q=\pi$ and $\omega=0$
and the intensity decays very fast with increasing energy for all $\vec{q}$.
Since this quantity or its value at $q=\pi$ and $\omega=0$ can be measured
by neutron scattering or NMR relaxation \cite{imai}, it is definitely necessary
to study the dependence on doping, temperature and $\Jast$ more systematically.
This is left for a future publication.
\section{Summary and outlook}
We presented a theory and results for the magnetic properties of the $t$-$J$
model in the framework of the dynamical mean-field theory, which treats
both the correlated hopping of the fermionic degrees of freedom and the
nonlocal exchange coupling between the spin degrees of freedom on the same
footing. As has been pointed out \cite{janis}, this approach ensures
a thermodynamically consistent description of the properties of the system
and especially does not introduce artificial phase transitions like e.g.\ in
slave-boson mean field theories.

One in our
opinion particularly interesting result is that the dynamical susceptibility
of the $t$-$J$ model can be expressed in an RPA-like fashion by the
susceptibility
of the Hubbard model at $U=\infty$ (cf.\ equation (\ref{chitJ_final})).
In addition the latter can be split into a local part plus a
$\vec{q}$-dependend
renormalization which for low doping turned out to be relatively small and only
moderately varying with $\vec{q}$. We find that in the case $\Jast=0$
(i.e.\ $U=\infty$) the absence of an explicit magnetic exchange
leads to $\chi^{HM}_{q=\pi}\approx\chi^{HM}_{loc}$ and an interesting
enhancement of the ferromagnetic correlations. This is in
contrast to the HM at finite $U$, where the effective magnetic exchange
$J\sim t^2/U$ leads to a strongly enhanced susceptibilty at $q=\pi$ and
a suppression at $q=0$ instead.
However, for the situation considered
here -- simple hypercubic lattice with nearest-neighbour hopping only -- we did
not observe a tendency towards a magnetic instability at $q=0$ for finite
doping, in accordance
with results obtained by other groups. The occurence of an enhanced
ferromagnetic susceptibility for Hubbard model in the limit $U=\infty$
nevertheless motivates a more detailed investigation of the mean-field
properties of the Hubbard model in this particular limit for different lattice
structures and
longer-range hopping.

A finite magnetic exchange $\Jast$ again strongly enhances the
antiferromagnetic susceptibility.
When one further increases $\Jast$ one eventually encounters a phase transition
into an antiferromagnetic phase
at a critical
value $\Jast_c(T,\delta)$.
 From our results of $\chi_{AF}^{HM}(T,\delta)$ we extracted the phase diagram
$T_N(\delta,\Jast)$. We found that $T_N$  increases monotonically as function
of $\Jast$ and -- for fixed $\Jast$ -- decreases monotonically as function of
$\delta$. For larger doping $\delta$ we observed that the curves
$T_N(\delta)$ for different but fixed values of $\Jast$ are almost
linear. This linearity agrees at least qualitatively with DMFT results for
the Hubbard model at finite $U$, where one finds a crossover from
standard weak-coupling behaviour in $T_N(\delta)$ for small $U$ to an almost
linear variation for $U\ge7\tast$. However, in contrast to our results one
observes a much faster depression of $T_N$ as function of $\delta$ and in
addition a transition into an incommensurate phase for large $\delta$.
Especially
the latter feature was not reproduced in our calculations.
The linearity of $T_N(\delta)$ finally allowed us to extrapolate our data
to obtain an approximation for the magnetic phase boundary of the $t$-$J$ model
at $T=0$.

The relatively weak dependence of the susceptibility of the HM on $\vec{q}$
was used to set up an approximation for the dynamical susceptibility by
assuming $\chi^{HM}_{\vec{q}}(\omega)\approx\chi^{HM}_{loc}(\omega)$, thus
giving to some extent a microscopic justification of the results in reference
\cite{scalapino}. Since in addition
$\chi^{HM}_{loc}(\omega)$ can be calculated fairly easy from the effective
single-site problem we were able to present results for the dynamical
spin structure factor $S^{tJ}_{\vec{q}}(\omega)$. The general expected
features, i.e.\ sharp maximum at $q=\pi$ and $\omega=0$, a shift of the maximum
to finite $\omega$ for $q<\pi$ and a fast decay as $\omega>0$, are well
reproduced.
There are of course several questions left. First of all one should
check the assumption of a nearly $\vec{q}$-independent
$\chi^{HM}_{\vec{q}}(\omega)$
carefully for several values of doping and temperature. Second a systematic
study
of $S^{tJ}_{\vec{q}}(\omega)$ as function of doping and temperature is clearly
needed. Another important issue not yet addressed concerns phase separation in
the $t$-$J$ model, which among other problems requires e.g.\ the evaluation of
the compressibility in
the antiferromagnetic phase. Work along this line is in progress.

Finally, one should stress again that the results presented here were
calculated
with a generalized mean-field theory or equivalently for the limit $d=\infty$.
This obviously means that their applicability to e.g.\ the $t$-$J$ model in
$d=2$ or $d=3$ is unclear. From high-temperature
expansions or exact diagonalizations for $d=2$ one knows for example that the
static homogenous susceptibility shows a nonmonotonic behaviour as function of
$\delta$, which may be attributed to fluctuations induced by the spin-flip term
in the model (\ref{tjm}).
Since the DMFT neglects this type of processes it is not too surprising that
in our results we always observe a monotonic decrease instead.
We thus do expect that the predictions of the DMFT will be modified not only
quantitatively but most likely also qualitatively, especially for
two-dimensional
systems.

\noindent{\bf Acknowledgements}:
We like to acknowledge useful discussions with
D.\ Vollhardt,
M.\ Jarrell,
W.\ Metzner,
F.\ Gebhardt,
P.\ van Dongen,
G.\ Uhrig,
K.\ Becker,
N.\ Grewe,
F.\ Anders,
and many others.
This work was supported by
the Deutsche Forschungsgemeinschaft grant number \mbox{Pr 298/3-1}.

One of us (TP) also wants to acknowledge the hospitality of the department
of physics at the University of Cincinnati, where part of this work was
done.
%

\end{document}